\documentclass[aps,preprint,showkeys,superscriptaddress,nofootinbib,amsmath,amssymb]{revtex4-1}
\usepackage{graphicx}
\usepackage{dcolumn}
\usepackage{tabularx}
\usepackage{bm}
\usepackage{epsfig}
\usepackage{color}

\begin{document}

\title{Unravelling the interplay of geometrical, magnetic and electronic properties of metal-doped graphene nanomeshes}

\author{Mohamed M.\ Fadlallah}
\email{mohamed.fadlallah@physik.uni-augsburg.de}
\affiliation{Institute of Physics, University of Augsburg, 86135 Augsburg, Germany}
\affiliation{Center for Fundamental Physics, Zewail City of Science and Technology, Giza 12588, Egypt}
\affiliation{Physics Department, Faculty of Science, Benha University, Benha, Egypt}

\author{Ahmed A. Maarouf}
\affiliation{Center for Fundamental Physics, Zewail City of Science and Technology, Giza 12588, Egypt}
\affiliation{Egypt Nanotechnology Center \& Department of Physics, Faculty of Science, Cairo University, Giza 12613, Egypt}

\author{Udo Schwingenschl\"ogl}
\email{udo.schwingenschlogl@kaust.edu.sa}
\affiliation{King Abdullah University of Science and Technology (KAUST), Physical Science 
and Engineering Division (PSE), Thuwal 23955-6900, Saudi Arabia}

\author{Ulrich Eckern}\
\affiliation{Institute of Physics, University of Augsburg, 86135 Augsburg, Germany}


\begin{abstract}
Graphene nanomeshes (GNMs), formed by creating a superlattice of pores in graphene, possess rich
physical and chemical properties. Many of these properties are determined by the pore geometry.
In this work, we use first principles calculations to study the magnetic and electronic
properties of metal-doped nitrogen-passivated GNMs. We find that the magnetic behaviour is
dependent on the pore shape (trigonal vs.\ hexagonal) as dictated by the number of covalent
bonds formed between the 3$d$ metal and the passivating N atoms. We also find that Cr and V doped
trigonal-pore GNMs, and Ti doped GNMs are the most favourable for spintronic applications. The
calculated magnetic properties of Fe-doped GNMs compare well with recent experimental
observations. The studied systems are useful as spin filters and chemical sensors.
\end{abstract}



\maketitle

\section{Introduction}

Two-dimensional (2D) graphene can be considered the basis of different graphetic materials: fullerene, nanotubes and graphite. 
Due to promising applications, graphene has been the subject to intensive research recently, experimentally as well as theoretically
\cite{Geim,Novoselov2011,CastroNeto2009,Rozhkov}. Pristine graphene has a zero band gap which restricts 
its  electro-optical applications. By reducing its dimensionality, quantum confinement results in a finite energy gap, as in quantum dots and nanoribbons \cite{Rozhkov}.
In such structures the band gap depends on the geometry, size, and edge passivation, hence the band gap can be tuned and thus 
the systems can be used in electro-optical technology \cite{DIS,MD}. 

Another graphene-based structure with a finite electronic band gap is a graphene nanomesh (GNM) \cite{Oswald2012,Petersen2011},
a 2D system formed by creating a lattice of pores in a graphene sheet. GNMs have been fabricated with pore sizes between 5 and 20 nm \cite{JB}. 
Stable doping of GNMs can be achieved by controlled passivation of the pore edges 
followed by the chelation of donor or acceptor atoms \cite{Maarouf2013}. GNM edges have also been used for gas separation \cite{DJ}, catalysis \cite{OA},
sensing \cite{RKP}, and hydrogen storage \cite{AD,SH}. Crown ethers have been suggested as a new route towards chemically functionalized graphene \cite{Guo2014}. 
For example, GNM nanostructures have the potential to become next-generation spintronic devices \cite{HXY,TKa,Lin2015}. 

The recent theoretical studies of graphene nanomeshes \cite{Oswald2012,Scheffler2006, Krashen,Santos2010,Maarouf2013,DJ,AD,SH,Guo2014,HXY}
and, in particular, the advances in experimental preparation of such systems \cite{JB,OA,RKP,Guo2014,TKa,Lin2015}
clearly suggest the need for further detailed investigations, in order to elucidate the structures with highest
application potential. In this context, we study the magnetic and electronic structure of
a dense regular array of pores in graphene, each of the pores being passivated by nitrogen, and doped with a 3$d$ transition
metal atom. While pore passivation by H and O also has been studied \cite{SH,Guo2014,TKa,Maarouf2013,Lin2015},
our focus here is on N (see also \cite{Lin2015}), which is known to form a stable
C--N--metal bond, for example, in metal phthalocyanine \cite{Fadlallah2016}. Two pore shapes of high symmetry are addressed, namely a
trigonal and a hexagonal one, as these are expected to be more stable, and easier to realize.

In Sec.\ II, the computational method is presented, then the thermodynamic stability is addressed on the basis of the binding energy, as well as the magnetic 
moments of different structures (Sec.\ III). Our results for the electronic structures are discussed in Sec.\ IV. Finally, a summary is given in Sec.\ V.

\section{Methodology and optimized N--GNM structures}

Spin polarised density functional theory calculations are performed using the projector augmented-wave pseudopotentials in the Vienna \textit{ab initio}
Simulations Package \cite{GK,GK1}. 
For the exchange-correlation energy density functional, the generalized gradient approximation \cite{ADB,JP} in the scheme of Perdew-Burke-Ernzerhof \cite{JP1} is utilized to obtain the optimized structures. 
Projector augmented-wave pseudopotentials for the valence electrons are employed. 
The wave functions are expanded in plane waves up to a cutoff energy of $600$ eV. 
A Monkhorst-Pack $k$-point mesh \cite{HJM} of $12$ $\times$ $12$ $\times$ $1$ 
is used for geometry optimisation, for which we employ the standard conjugate-gradient method,
until the largest force on the atoms becomes smaller than $0.01$ eV/{\AA}, and the tolerance of total energy reaches $10^{-6}$ eV.
As initial position for the optimisation process, the dopant atom is placed above the centre of the respective pore, 
at a height of 4 {\AA}.
The nanomeshes are modeled as layered three-dimensional systems, with a vacuum of $15$ {\AA} in z--direction between the individual layers, in order to
avoid any layer-layer interaction. The supercell dimensions in the x--y--plane, chosen to be $6$ $\times$ $6$ in this work, are kept fixed for all calculations.

In our study, we consider two GNM configurations. A trigonal ({\it t}) pore is formed by removing $6$ C atoms, then passivating the pore edge by $3$ N atoms; 
we denote this configuration by {\it t}-N--GNM. A hexagonal ({\it h}) pore configuration is formed by removing $12$ C atoms, passivating the pore edge with $6$ N atoms, denoted as {\it h}-N--GNM. Then $3d$ transition metals are placed in the pore, see Fig.\ 1, and their effect on the N--GNM magnetic and electronic properties is studied.

\begin{figure}[h]
\includegraphics[width=0.25\textwidth]{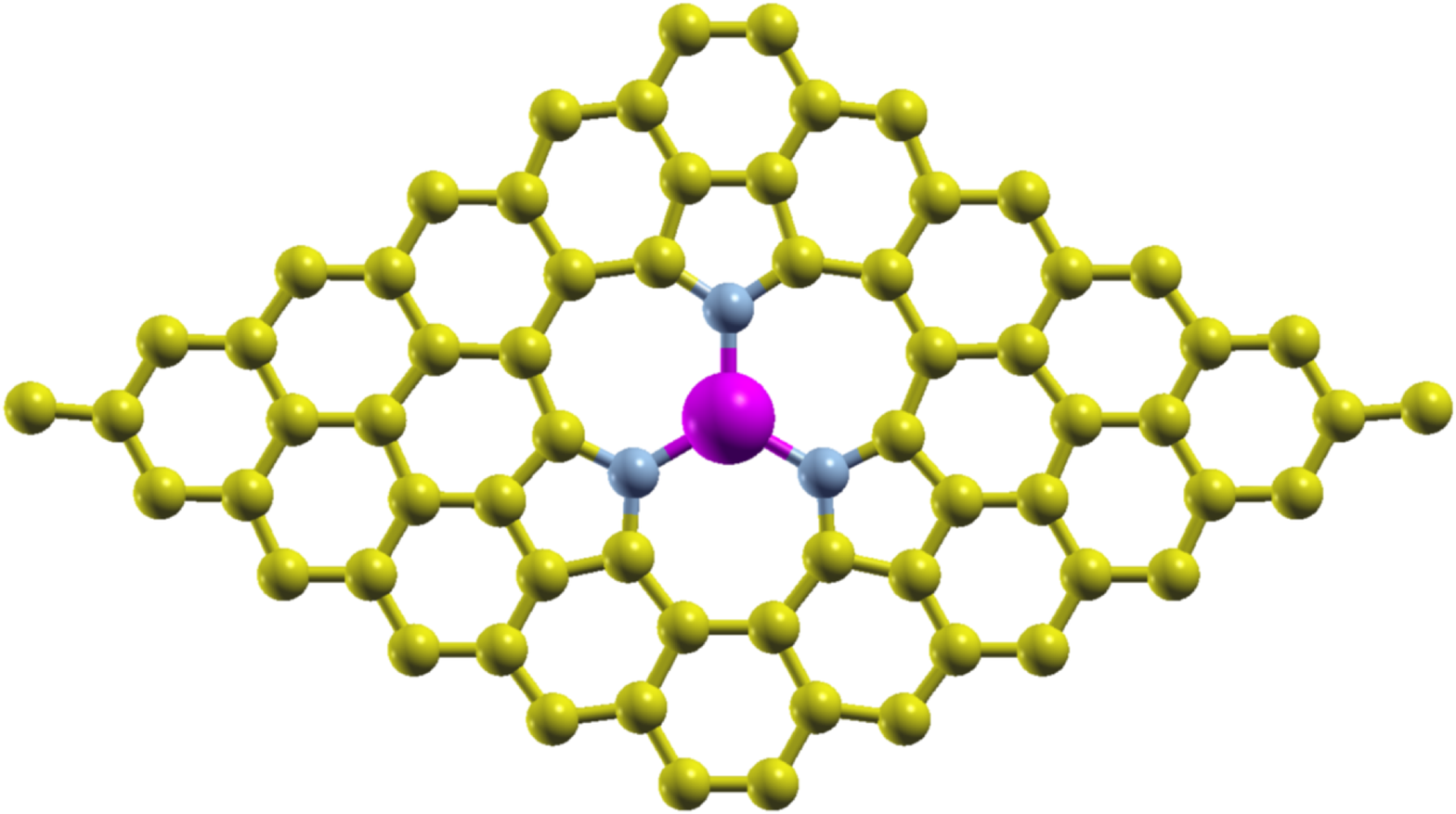}~\hspace{5mm}
\includegraphics[width=0.25\textwidth]{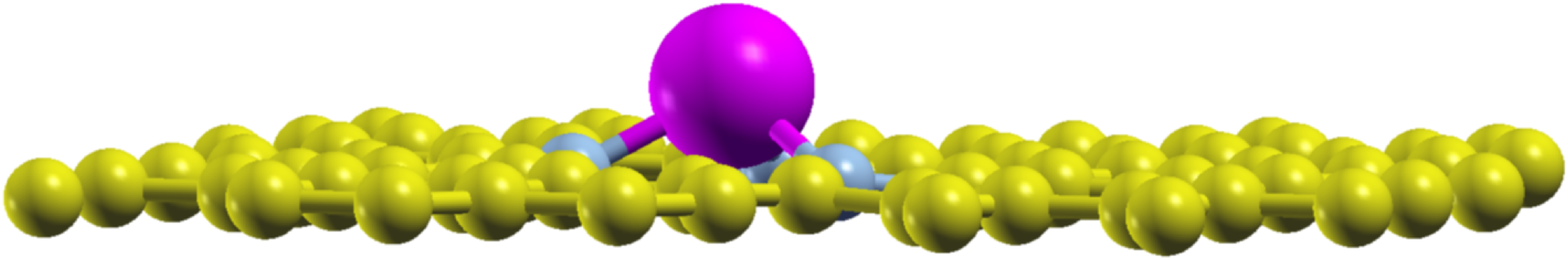}~\hspace{5mm}
\includegraphics[width=0.25\textwidth]{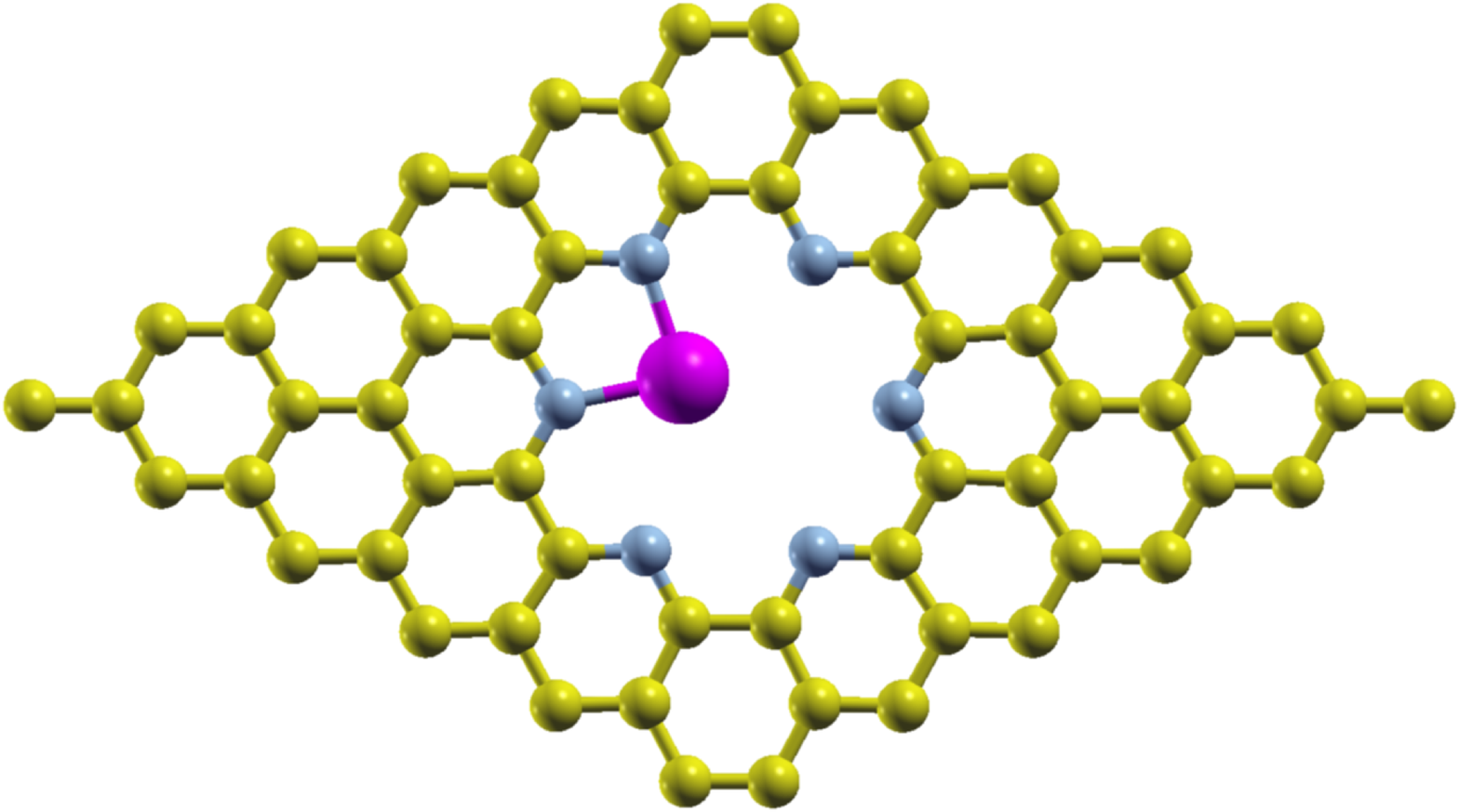}\\[5mm]
\caption{(Color online) Optimal configuration for metal-doped N--GNM: (from left to right) trigonal ({\it t}) pore (top view), {\it t}-pore (side view), and hexagonal ({\it h}) pore (top view);
 golden, grey, and pink spheres represent carbon, nitrogen, and metal atoms, respectively.}
\label{fig1}
\end{figure}

Since we are considering rather light metals in this work, we expect spin-orbit coupling to be of minor importance, as its effect 
increases with the fourth power of the atomic number. However, one should keep in mind that the spin-orbit interaction can, in principle,
have a distinct effect, especially for the states around the Fermi level, in certain `critical' cases, e.g., for the half 
metal system MnBi \cite{Mavropoulos2004}. In other cases, like NiMnSb, the modification is found to be very small, 
below 1\% \cite{Mavropoulos2004}. Some discussion of relativistic effects can also be found in a recent review paper \cite{Santos2010}.

\section{Stability and magnetic moment}

In the optimized structure, {\it t}- and {\it h}-N--GNMs have zero magnetic moments. Table I shows the bond lengths between metal and N atoms. 
The bond length decreases as the ionic size decreases from Sc to Cr in both pore configurations, while from Fe to Zn it is almost constant.
A slight increase of the bond length is found for Mn (see below).
The optimized structure is planar for metal-doped N--GNM in the case of the {\it h}-pore. 
However, for the {\it t}-pore the metal atom is located out of plane of the N--GNM sheet (see Fig.\ 1, side view) except for Zn. 
The vertical distance between the metal and the sheet plane decreases as the ionic size decreases, from $1.1$ {\AA} for Sc to $0.5$ {\AA} for Cu.

The binding energy ($E_{b}$), Table I, is an indicator of the stability of the respective structure; it can be calculated using the following equation:
\begin{equation}
E_{b}=E_{\textrm{Metal--N--GNM}}-E_{\textrm{N--GNM}}- E_{\textrm{Metal}},
\end{equation}
where $E_{\textrm{Metal--N--GNM}}$, $E_{\textrm{N--GNM}}$, and $E_{\textrm{Metal}}$ are the energy of the metal-doped N--GNM, the energy of
the N--GNM, and the isolated metal in the same volume, respectively. 
The binding energy assesses the strength of the interaction between N--GNM and the transition metal atom. Roughly speaking, the 
binding strength ($\sim | E_b |$) increases when considering the series from Zn to Sc, i.e., from right to left in Table I (with the 
exception of Co in the $t$-pore). This is consistent with the fact that along this series the number of empty outer-shell electron
states on the metal increases, allowing for increasing hybridisation with N, and thus stronger binding. It is not surprising that
the metals in the {\it t}-pore are more strongly bound than the corresponding ones in the {\it h}-pore, since there are three
bonds with N in the former and only two bonds with N in the latter case. One should note, however, that stronger binding corresponds
to longer bond lengths: the latter is consistent with the increase of the covalent radius from Zn $\dots$ Sc \cite{Cordero2008}.

The magnetic moment, see Table I, along the sequence from Sc to Cr, with open $d$ shells on the metal, is related to the 
number of remaining electrons after forming the metal--N bonds. Two and three electrons are shared with the N atoms in
the {\it h}- and the {\it t}-pore, respectively, as shown in Fig.\ 1.  
Therefore the magnetic moment of the metal-doped {\it h}-N--GNM is larger than that of the {\it t}-N--GNM doped with the same metal by one. 
For Mn which is in the high-spin state only two electrons engage with the N atoms in both cases, {\it t}- and {\it h}-pore. 
Note that Mn in the high-spin state is known to have a larger covalent radius than in the low-spin state \cite{Cordero2008}, 
in agreement with the higher Mn--N bond length in comparison with its `neighbour' elements.

For Fe in the {\it t}-pore apparently all un-paired $d$ electrons bond with the N atoms, thus the magnetic moment becomes zero. This
agrees with the recent experimental observation \cite{Lin2015} that Fe assumes a low-spin state if surrounded by a large number of N
atoms. The spin state of Fe changes, however, when Fe is embedded in the {\it h}-pore where only two electrons are shared with the N atom.
Thus Fe in the {\it h}-pore appears to be similar to Fe in a double vacancy \cite{Lin2015}. We note that the magnetic moments in the 
sequence Sc $\dots$ Fe are larger for the {\it h}-pore that for the {\it t}-pore, which is consistent with general crystal field theory: a large
hole should support higher spin states as compared to a small one \cite{Krashen}.

For Co and Ni, the three and two un-paired electrons, respectively, bond with the N atoms in the {\it t}-pore. For these metals, which
are near closed shell, the magnetic moment for the {\it h}-pore is {\em smaller} than the magnetic moment of the corresponding doped {\it t}-pore,
which indicates a strong hybridisation of the outer shell electrons ($s$ and $d$ orbitals) with N. Accordingly, the bond lengths for Co and Ni
are about 7\% smaller than the bond length Fe--N.
Regarding closed atomic shells, Cu and Zn are non-magnetic in {\it t}- and {\it h}-pores to a good approximation. 

In comparison with previous work, we note that for the cases Sc, Fe, and Zn the
magnetic moments of metal-doped {\it t}-N--GNMs are very similar to the same metal substituted into pristine graphene \cite{Krashen}, 
and to graphene with a single vacancy \cite{Santos2010}. For metal-doped {\it h}-N--GNMs our results are very similar for V, Fe, Co, Cu, 
and Zn embedded in graphene with a double vacancy \cite{Krashen}. For the latter case, the metal--N bonds have the same effect on the 
magnetic moment as metal--C bonds. Apparently, for the control of the magnetic state it is important how many C atoms are removed,
how they are removed (which relates to the shape of the pores), and which passivating element is used.

\begin{center}
\begin{table}[htb]
\caption{Bond lengths (\AA), binding energies ($E_{b}$ (eV)), and magnetic moments ($\mu_{B}$), respectively, of 
metal-doped {\it t}- and {\it h}-N--GNM}
\begin{tabular}{|c|c|c|c|c|c|c|c|c|c|c|c|c|}
\hline\hline
 Metal& &Sc& Ti& V& Cr& Mn&Fe& Co& Ni& Cu& Zn
\\ [-0.5ex]
\hline
\hline
Configuration& & $3d^{1}4s^{2}$& $3d^{2}4s^{2}$& $3d^{3}4s^{2}$& $3d^{5}4s^{1}$& $3d^{5}4s^{2}$& $3d^{6}4s^{2}$& $3d^{7}4s^{2}$& $3d^{8}4s^{2}$& $3d^{10}4s^{1}$& $3d^{10}4s^{2}$
\\ [-0.5ex]
\hline
   \hline

& {\it t}& 1.98 & 1.90  & 1.88  & 1.85 &1.92& 1.82&1.81&1.82&1.82&1.80 \\[-1ex]
\raisebox{1.5ex}{Bond length}& {\it h}&2.22   &2.10 & 2.10  &  2.10  &2.14  & 2.10&1.98&1.96&2.00 &2.00\\[-0.5ex]

\hline
  \hline

& {\it t}& --10.0&--9.4& --8.5 &--6.6&--6.6 &--6.4 &--7.3&--6.5 &--5.3  &--4.7 \\[-1ex]

\raisebox{1.5ex}{$E_{b}$} & {\it h}
&--7.4&--6.8& --5.9 &--4.3&--4.3 &--4.5 &--4.5& --4.4&--3.3& --1.4  \\[-0.5ex]

\hline
  \hline
&{\it t}& 0.0&1.0& 2.0 &2.6 &4.8&0.0 &3.5 &2.0 &0.4 & 0.0 \\[-1ex]

\raisebox{1.5ex}{Magnetic moment} & {\it h}
&0.7& 2.0 &3.0&3.4&4.4 &3.8 & 1.6 &1.4 & 0.0& 0.2 \\[1ex]
\hline
       \hline

\end{tabular}
\end{table}
\end{center}

\section{Electronic structure}
The density of states (DOS) of pristine graphene, which has been known for a long time \cite{Wallace1947}, 
features a characteristic linear behaviour near the Fermi energy ($E_{F}$), $\sim|E-E_{F}|$. This behaviour 
is characteristically modified when passivated pores are created.
Figure\ 2 shows the DOS for the {\it t}-N--GNM and {\it h}-N--GNM structures. 
For the {\it t}-pore a broad band of states is created around the Fermi energy ($-0.2$ to $0.7$ eV), and the contribution of N states appears at $-0.9$, $-0.7$ eV, and around the Fermi energy ($-0.1$ to $0.2$ eV). As compared with pristine
graphene, {\it t}-N--GNM system is metallic, Fig.\ 2a.
For N--GNM with {\it h}-pore, the DOS shows semiconducting behaviour with a band gap of $0.8$ eV (see Fig.\ 2b). The N states have a significant participations in the energy range $-1.3$ to $-0.6$ eV, and a small contribution in the conduction band
($1.2$ to $1.5$ eV). Clearly the shape of the pore, i.e., the number of removed C atoms and the passivation with N, crucially
influences the density of states of the nanomesh.

The surprising (at least at first sight) observation is the metallic behaviour of {\it t}-N--GNM. However, the $t$-pore structure
involves three closely interacting N atoms in the pore; this structure p-dopes the graphene by acquiring three electrons from it, as is confirmed by
integrating the DOS from the (pristine) valence band edge to the Fermi energy. Hence the Fermi energy is moved compared to the pristine case, 
rendering the system metallic. From another point of view, we note that semiconducting behaviour is observed for O passivation, see Fig.\ 2c in Ref.\ \onlinecite{Maarouf2013}: starting from there and replacing O by N, the latter having three electrons less than the former, we arrive
at the same conclusion. Of course, these results depend, as is well known \cite{Petersen2009,Liu2009}, on the concrete size of the pore and the
supercell: for example, increasing the latter will generally decrease the effect of the pore, i.e., reduce the doping level, 
thereby shifting the Fermi energy towards the (pristine) valence band edge.

\begin{figure}[htb]
\includegraphics[width=0.60\textwidth]{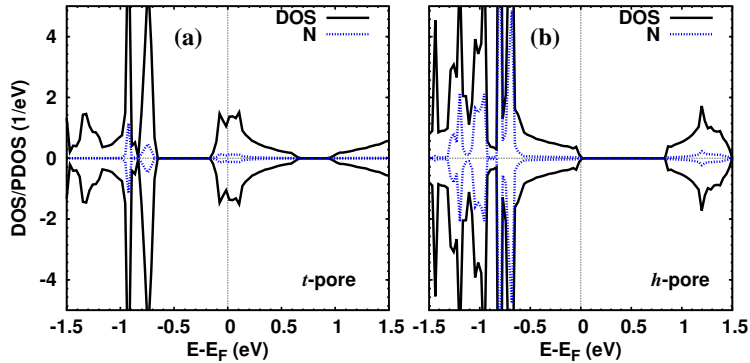}~\hspace{5mm}
\caption{(Color online) Density of states (DOS) and partial density of states (PDOS) of N atoms, for {\it t}-N--GNM (a) and {\it h}-N--GNM (b)}
\label{fig2}
\end{figure}
 
When adding Sc as a dopant to the {\it t}-pore, as compared with {\it t}-N--GNM (Fig.\ 2a), the system becomes a semiconductor with a band gap of $0.5$ eV, and the Fermi
energy shifts towards higher energy, see Fig.\ 3a. The Sc states appear only in the conduction band above $0.8$ eV, and the N states do not have a significant contribution in the DOS. The symmetry of spin up and spin down states results in a zero magnetic moment. 
With respect to Sc in the {\it h}-pore (Fig.\ 3b), the hybridisation between Sc and N is stronger than for Sc in 
the {\it t}-pore, and lies in a broad range (above $-0.3$ eV). This hybridisation creates covalent bonds between Sc and the N atoms.
As compared with the DOS of {\it h}-N--GNM (Fig.\ 2b), the Fermi energy moves to the conduction band, and the structure becomes metallic.
The asymmetry between spin up and spin down states leads to a magnetic moment of $0.7$ $\mu_{B}$.
\begin{figure}
\includegraphics[width=0.60\textwidth]{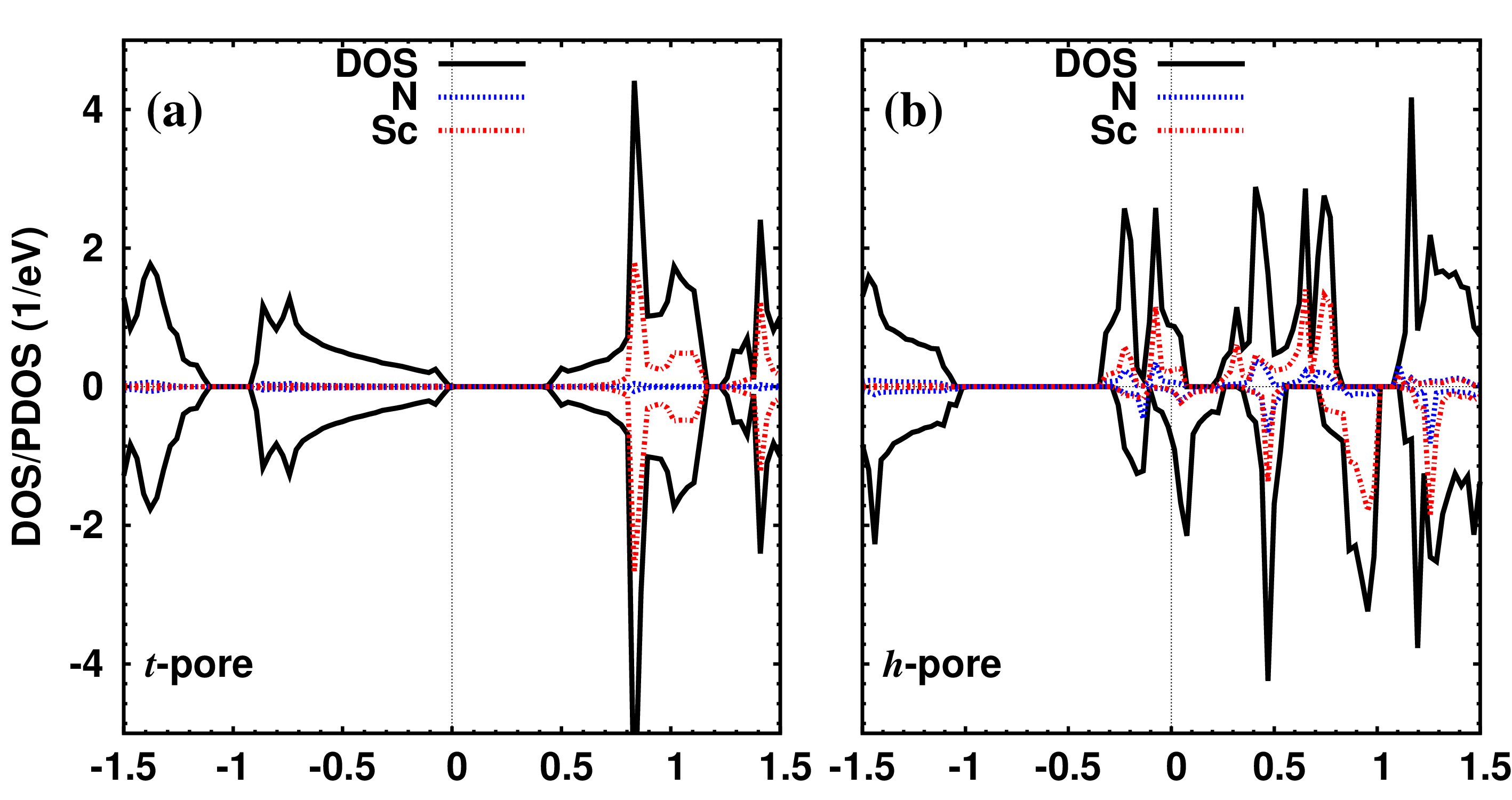}\
\includegraphics[width=0.60\textwidth]{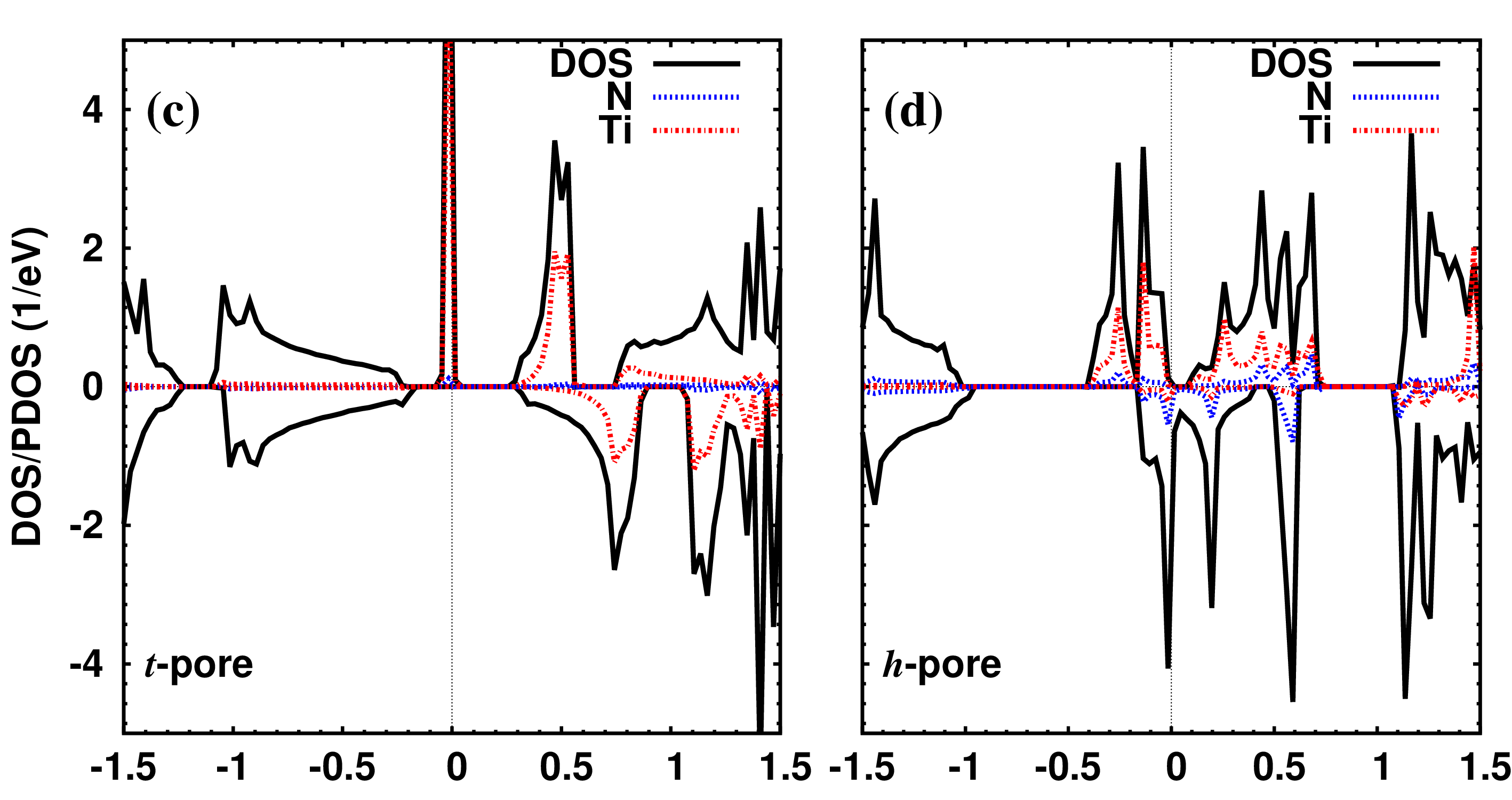}\
\includegraphics[width=0.60\textwidth]{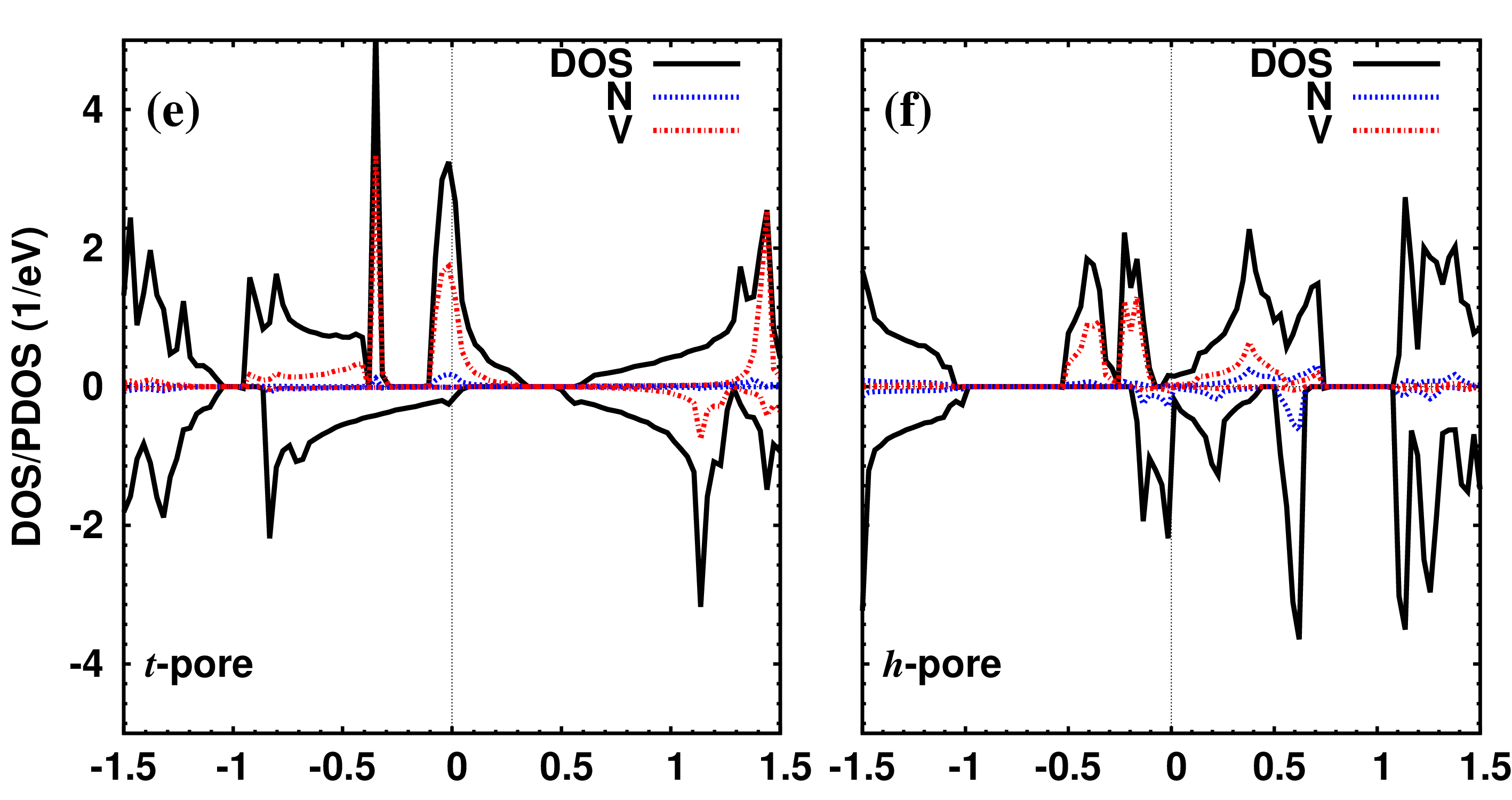}\
\includegraphics[width=0.60\textwidth]{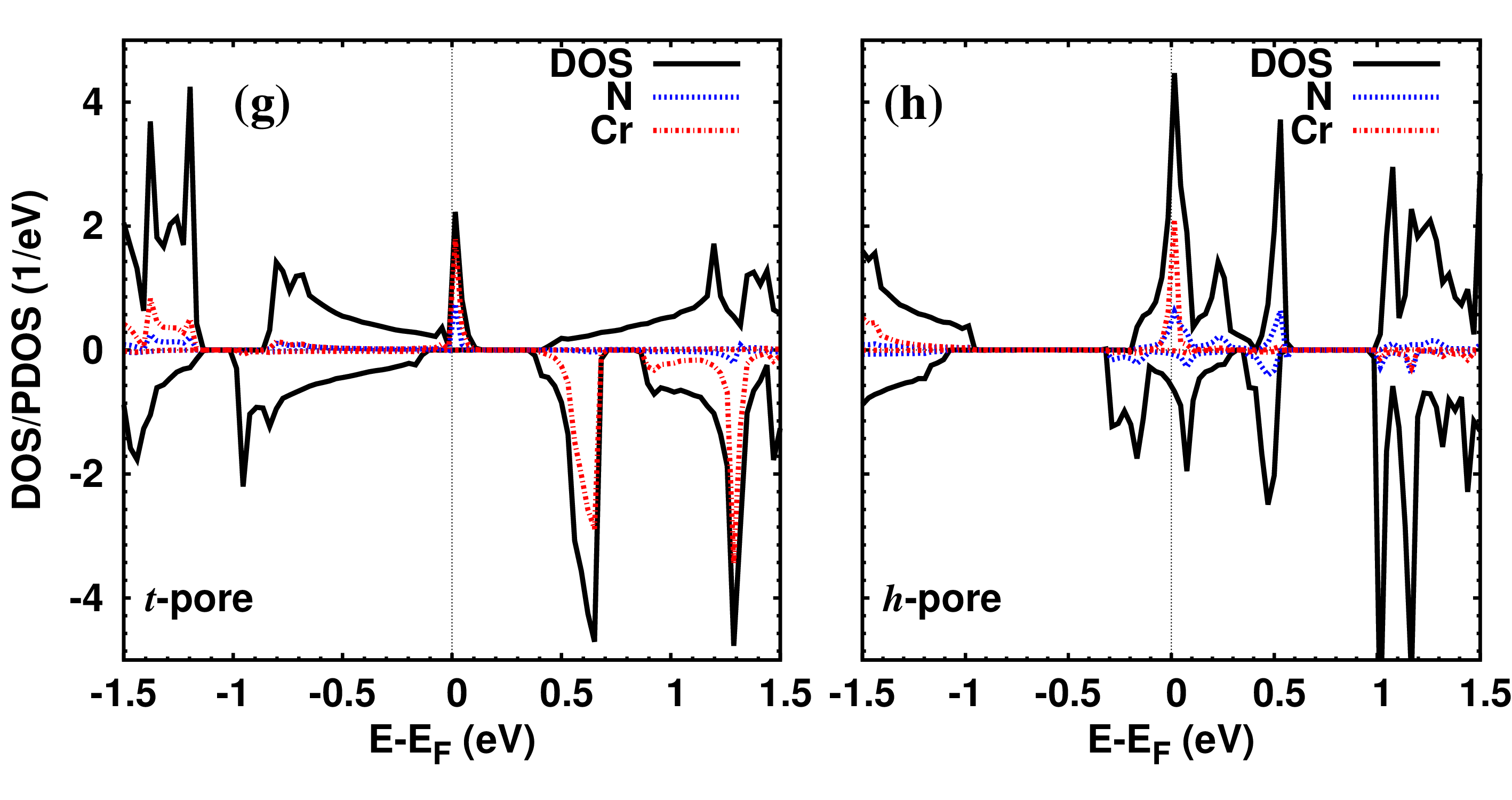}
\caption{(Color online) Density of states (DOS) and partial density of states (PDOS),
for Sc (a, b), Ti (c, d), V  (e, f), and Cr (g, h) doped N--GNM; {\it t}-pore (left) and {\it h}-pore (right)} 
\label{fig3}
\end{figure}

The density of states for a {\it t}-pore doped with Ti is shown in Fig.\ 3c, demonstrating the covalent bonds between Ti and N orbitals, in particular, at the Fermi energy. 
The Fermi energy is shifted further to higher energy (compared with the corresponding pore with Sc). 
A strong hybridisation between Ti and C states appears in the whole energy range of the conduction band. The spin up states at the Fermi energy originate from Ti states. Since the Ti atom has an additional $d$ electron compared with Sc,
the magnetic moment of this system is $1$ $\mu_{B}$, as we see also in the strong asymmetry of the two spin components.
This system may be useful for spintronic devices due to the metallic spin-up states of Ti at the Fermi energy. 
In the case of the hexagonal pore with a Ti atom (Fig.\ 3d), the Fermi energy is shifted towards higher energy as compared with {\it h}-N--GNM, and covalent bonds due to the hybridisation
between Ti and N states are formed. The $3d$ Ti states for spin up are located around the Fermi energy in the range $-0.5$ to $0.7$ eV.
The asymmetry between spin up and spin down components is higher than for Ti in the {\it t}-pore, so the magnetic moment
becomes $2$ $\mu_{B}$. Due to the asymmetry of two spin components, this system can be used as spintronic device.

Figure 3e shows the density of states for V inside the {\it t}-N--GNM. A strong overlap between V and C states is seen in the valence and the conduction bands. In particular, there is 
a distinct overlap at the Fermi energy with a broad peak, showing that the metal and N atoms have a significant contribution in one spin component. 
As compared with {\it t}-N--GNM (Fig.\ 2a), the contribution of V states leads to a splitting of the two spin components. 
For the {\it h}-pore containing V, as compared with {\it h}-N--GNM, the Fermi energy is shifted towards higher energy as shown in Fig.\ 3f. 
Due to the contributions of V states in one spin direction (in the energy range $-0.6$ to $0.8$ eV), there is a clear spin splitting in the DOS in this range. 
There is no distinct contribution of V states at the Fermi energy, compared with V in {\it t}-N--GNM, and the N states appear in the energy $-0.2$ to $0.8$ eV. 

Now filling the {\it t}-pore with a Cr atom, Fig.\ 3g, the DOS is similar to the V 
doped {\it t}-N--GNM at the Fermi energy. 
The contribution of Cr states at the Fermi energy creates a clear spin splitting. Also, the Cr and N states
contributions are apparent for one spin direction. The participation of Cr states is pronounced in the other spin component of the conduction band. 
Adding Cr in the {\it h}-pore, Fig.\ 3h, a distinct peak contribution of Cr and N states appears at the Fermi energy for one spin. 
Due to the asymmetry of the two spin orientations, this system can be used as spin filter device.

\begin{figure}
\includegraphics[width=0.60\textwidth]{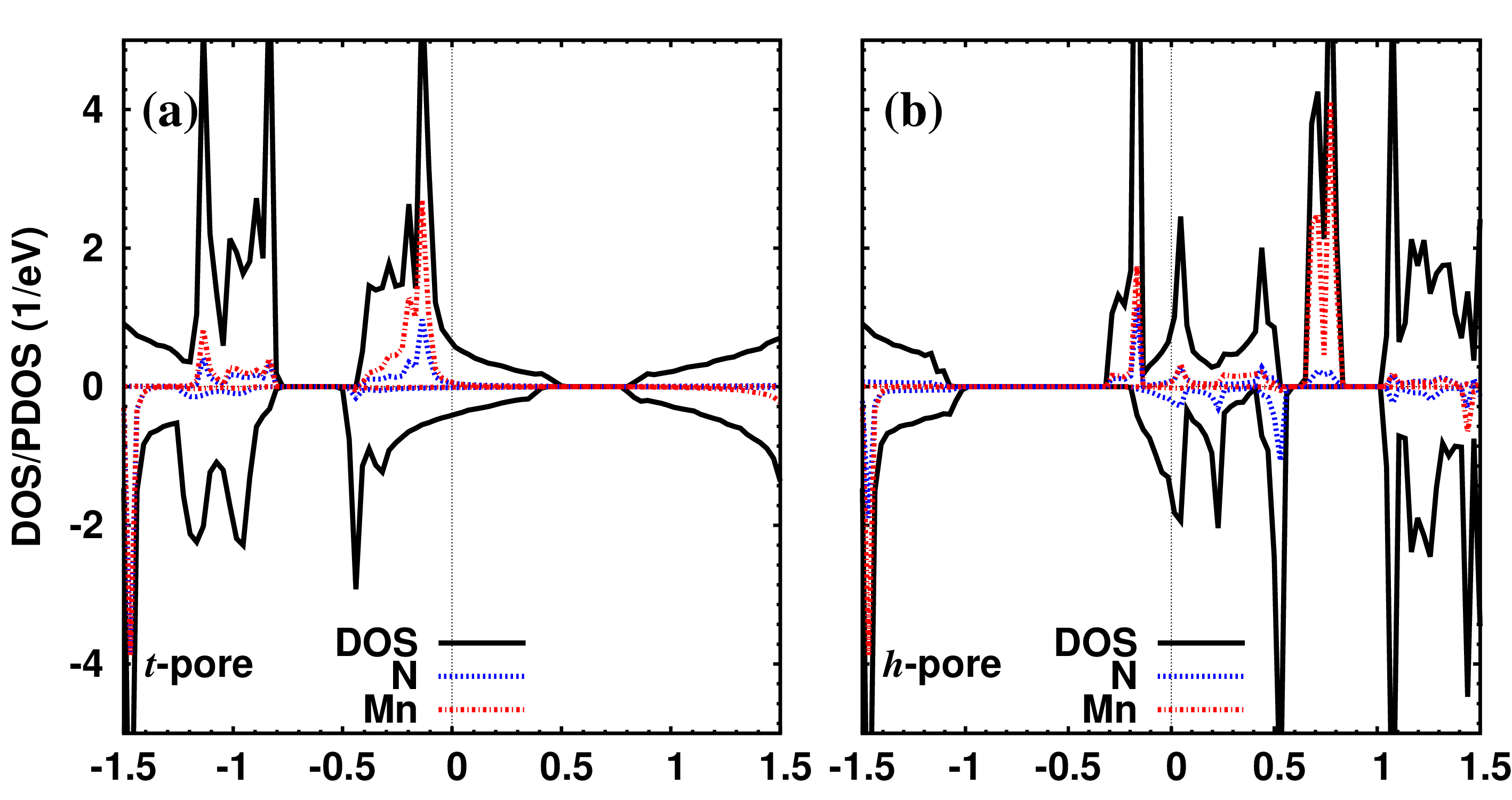}\
\includegraphics[width=0.60\textwidth]{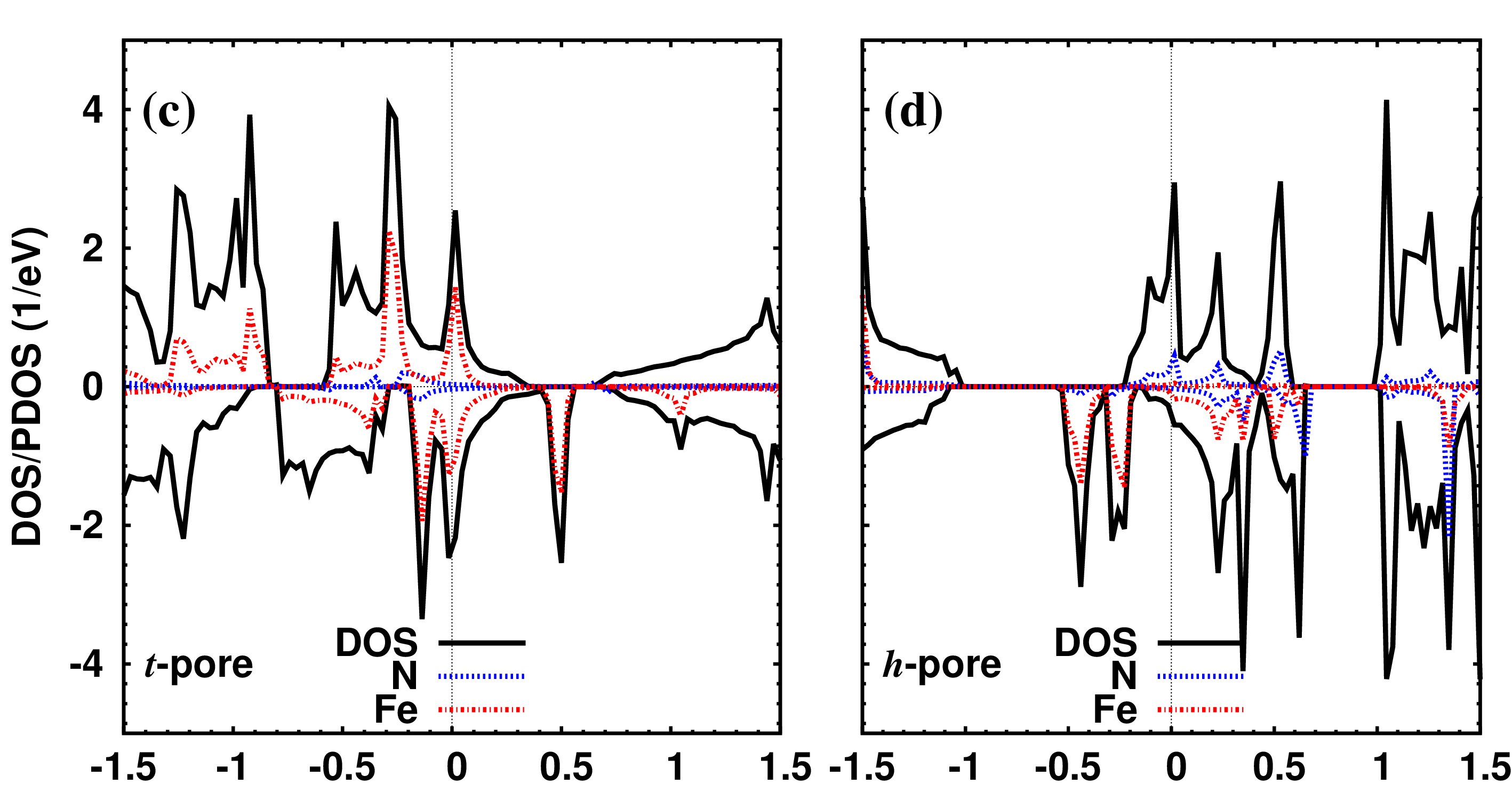}\
\includegraphics[width=0.60\textwidth]{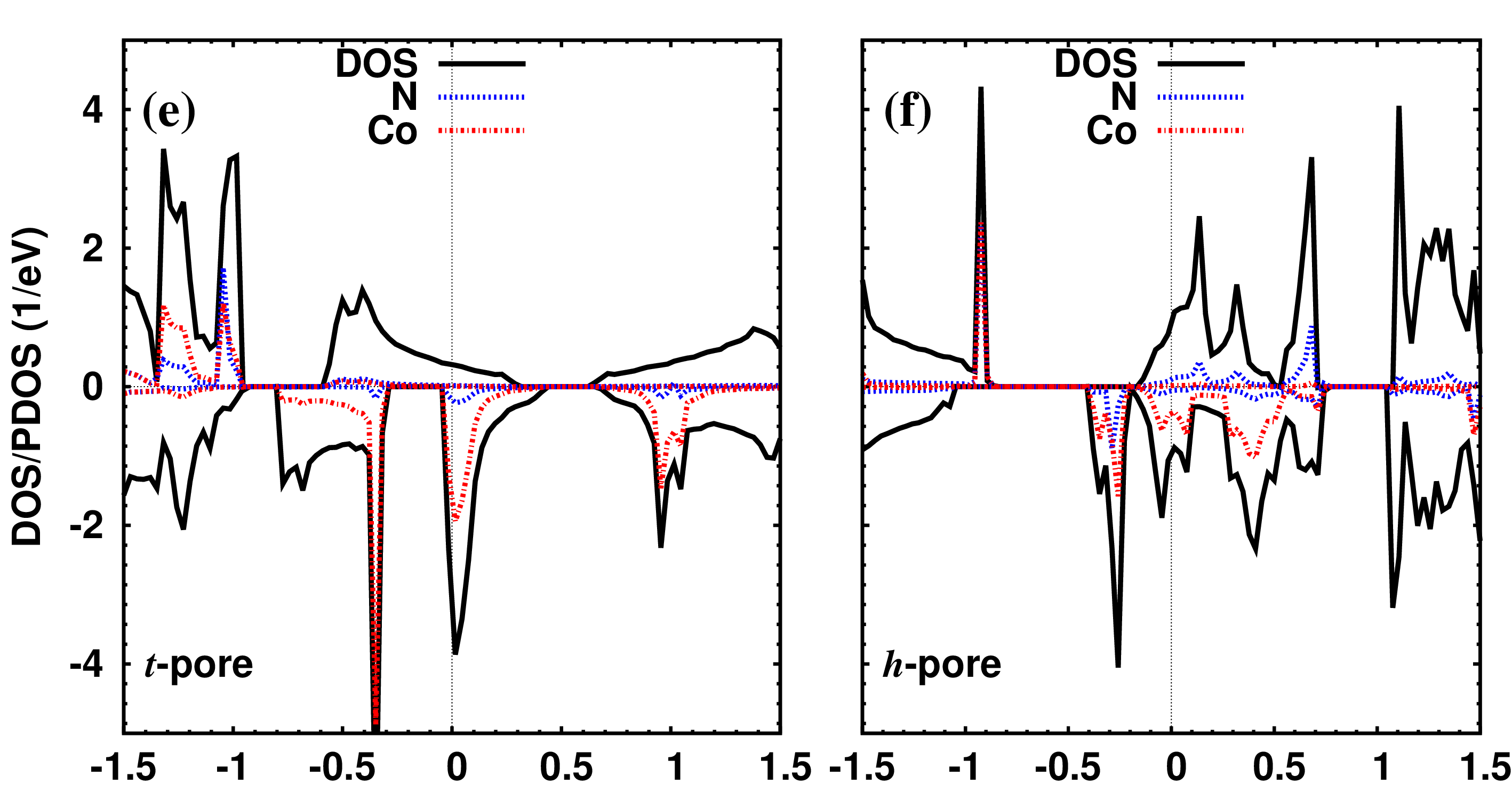}\
\includegraphics[width=0.60\textwidth]{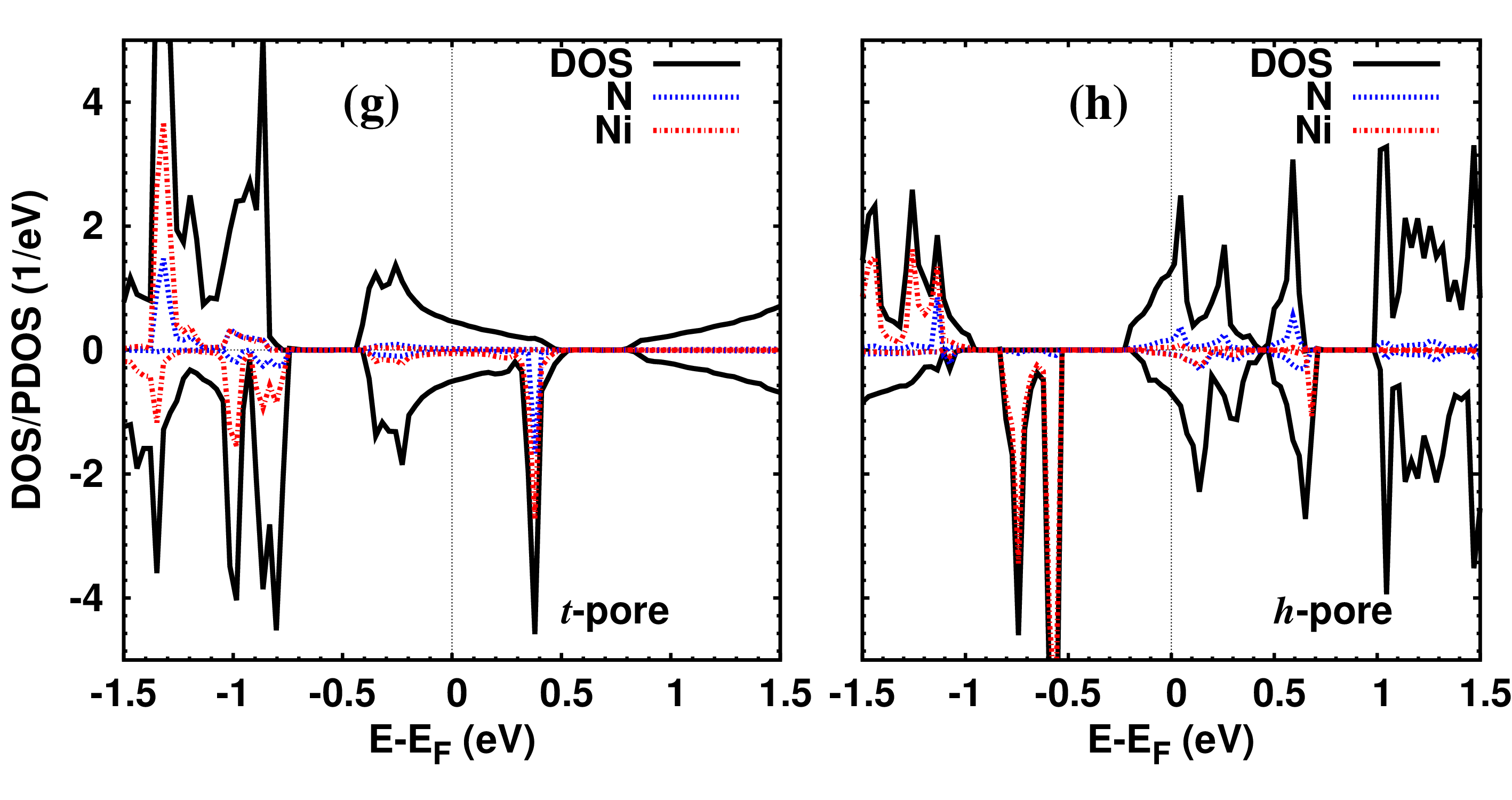}
\caption{(Color online) Density of states (DOS) and partial density of states (PDOS), 
for Mn (a, b), Fe (c, d), Co (e, f), and Ni (g, h) doped N--GNM; {\it t}-pore (left) and {\it h}-pore (right)} 
\label{fig4}
\end{figure}

The density of states of Mn in the {\it t}-N--GNM is illustrated in Fig.\ 4a. 
The Mn states are located in the valence band, and a small splitting in spin components around the Fermi energy is observable. 
The bonding between N and Mn orbitals appears at $-1.5$ eV, and in the energy ranges $-1.1$ to $-0.8$ eV, and $-0.4$ to $-0.1$ eV. 
The Mn dopant does not change the metallic states of {\it t}-N--GNM. 
For the {\it h}-pore doped with Mn, the Mn states contribute to the valence and the conduction bands (spin up only). 
A small contribution from Mn near the Fermi energy can be noticed. Some hybridisation between Mn and N states is seen in the whole energy range, see Fig.\ 4b. 
The structure is metallic, in contrast to undoped N--GNM with {\it h}-pore.

The effect of Fe on the DOS of {\it t}-N--GNM system is addressed in Fig.\ 4c. The Fe states span 
the energy range  from $-1.5$ to  $0.2$ eV, and have a significant contribution near $0.5$ eV and $1$ eV 
for spin down. The N states are located in a small energy range from $-0.4$ to $-0.1$ eV.  
As in most cases for the {\it t}-pore, this system is metallic. For the {\it h}-pore with Fe, the metal states have a contribution in one spin direction only in the energy 
range $-0.5$ to $1.4$ eV, with notable contributions from N states. In the same energy range for the other spin component, the N states contribute significantly. 
The splitting in the two spin components around the Fermi energy from $-0.5$ to $0.7$ eV is apparent, see Fig.\ 4d. 
Therefore the magnetic moment in this case is larger than that of Fe in the {\it t}-pore. 
Hence an Fe doped {\it h}-N--GNM is a promising system for spin filter applications. 

When a Co atom is placed in a {\it t}-pore, see Fig.\ 4e, the electronic structure is very similar to that of V, Fig.\ 3e. However, there are 
stronger contributions of Co states around $-1.2$ eV which enhance the magnetic moment to $3.5$ $\mu_{B}$, as compared with the V system.
The contribution of Co states inside {\it h}-N--GNM in one spin direction is higher than in the other, see Fig.\ 4f. The contribution of N states in the energy
range $-0.5$ to $0.7$ eV is similar to that of N in the {\it h}-pore with Fe (Fig.\ 4d). Metallic states appear at the Fermi energy, with 
a splitting between the two spin orientations due to the contribution of Co states.

Turning to closed $3d$ shell atoms (Ni, Cu, and Zn), we show the DOS for Ni doping only, see Figs.\ 4g and 4h.
In the {\it t}-pore the Ni states contribute to the valence band, and at $0.4$ eV, Fig.\ 4g.
The shift of the Fermi energy is smaller than in previous structures. The main contributions of the Ni atom in the {\it h}-pore, Fig.\ 4h, are found 
in the valence band, and only a little in the conduction band. For Cu in the {\it t}-pore, the Cu states are found below 
$0.1$ eV, and there is approximate symmetry in the two spin components, therefore the magnetic moment is small, $0.4$ $\mu_{B}$. 
The spin symmetry also appears for Cu inside the {\it h}-pore, where the Cu states are located in the valence band. 
For Zn in the {\it t}-pore, there is only a shift of the Fermi energy towards higher energy, as compared with undoped {\it t}-N--GNM, 
without any contributions from Zn states, so the system remains non-magnetic. 
For the {\it h}-pore, the Zn states contribute to the DOS at $0.3$ eV, but approximate spin symmetry is observed; 
hence again the magnetic moment is small, $0.2$ $\mu_{B}$.

\begin{table*}[htb]
\caption{Spin polarisation ratio in $\%$}
\centering
{
\begin{tabular}{|c|c|c|c|c|c|c|c|c|c|c|}
\hline
  \hline
Metal&Sc& Ti& V& Cr& Mn&Fe& Co& Ni& Cu& Zn\\  [-0.5ex]
\hline
 \hline                        
{\it t}-N--GNM

&0&100 & 91 &100 &23&0&77  &61&7&0    \\  [-0.5ex]
\hline
  \hline

{\it h}-N--GNM

\raisebox{2ex}
&19&100 &87 &66&25&72&6&27&0 &77 \\ [1ex]
\hline
  \hline
\end{tabular}
}
\label{table2}
\end{table*}
 
For spintronic and spin filter devices, the spin resolved density of states at the Fermi energy is decisive. Hence we finally
determine the spin polarisation ratio ($P_{\rm DOS}$) at the Fermi level, $E_{F}$: 
\begin{equation}
\label{eq1}
P_{\rm DOS} = \frac{D_{\uparrow}(E_{F})-D_{\downarrow}(E_{F})}{D_{\uparrow}(E_{F})+D_{\downarrow}(E_{F})} ,
\end{equation}
where $D_{\uparrow}$ and $D_{\downarrow}$ denote the density of states of the spin up and spin down states, 
respectively \cite{Kokado2006,Kokado2016}.\footnote{The `spin polarisation ratio' is, of course, an often used concept, 
and it can refer to different quantities like the particle number, the density of states, the resistance, and others. 
Here, it is the `DOS' version which is appropriate in view of transport applications. See, e.g., Refs.\ \onlinecite{Kokado2006, Kokado2016}.}
The results are given in Table II. We conclude that the systems with Ti in both pores, and Cr in the {\it t}-pore can be used as spintronic devices.
Furthermore, we can use V in both pores, Fe and Zn in the {\it h}-pore, and Co in the {\it t}-pore for spin filter applications.
On the other hand, Sc and Cu doped N--GNMs have only a low spin filter efficiency.

\section{Summary}

Using spin-polarised density functional theory, we find that the state of semi-metallic graphene is altered to metal and semiconductor in a 
graphene nanomesh passivated by N (N--GNM) with trigonal and hexagonal pores, respectively. The covalent bonds between N and the $3d$ metal are 
the basis for understanding the magnetic
and electronic properties of metal-doped N--GNM. A splitting in the density of states of the two spin components is created due to the contributions
of the $3d$ metals, except for Sc and Zn in the trigonal pore. We find doped N--GNMs to be magnetic for most $3d$ metals inside the trigonal pore
(i.e., except for 
Sc, Fe, and Zn). In the case of the hexagonal pore, all doped N--GNMs are magnetic except for Cu. A 100\% spin polarisation at the Fermi level
appears when Ti is used as a dopant in both pores (trigonal and hexagonal), as well as for Cr in the trigonal pore. These structures will produce 
a pure spin current, which can be used in spintronics and nanoelectronics devices. In addition, we find V inside both pores, Co in the trigonal 
pore, and Zn in the hexagonal pore, to be useful for spin filter and chemical sensor applications.

\acknowledgments{We acknowledge financial support by the Deutsche Forschungsgemeinschaft (through TRR 80). The research
reported in this publication was supported by funding from King Abdullah University of Science and Technology (KAUST).}


\begin{thebibliography}{10}

\bibitem{Geim} A.\ K. Geim, Science {\bf 324}, 1530 (2009).
\bibitem{Novoselov2011} K.\ S.\ Novoselov, Rev.\ Mod.\ Phys.\ \textbf{83}, 837 (2011).
\bibitem{CastroNeto2009} A.\ H.\ Castro Neto, F.\ Guinea, N.\ M.\ R.\ Peres, K.\ S.\ Novoselov, and A.\ K.\ Geim, Rev.\ Mod.\ Phys.\ \textbf{81}, 109 (2009).
\bibitem{Rozhkov} A.\ V.\ Rozhkov, G.\ Giavaras, Y.\ P.\ Bliokh, V.\ Freilikher, and F.\ Nori, Phys. Rep. {\bf 503}, 77 (2011).
\bibitem{DIS} D.\ I.\ Son, B.\ W.\ Kwon, D.\ H.\ Park, W.\ S.\ Seo, Y.\ Yi, B.\ Angadi, C.\ L.\ Lee, and W.\ K.\ Choi, Nat.\ Nanotechnol.\ {\bf 7}, 465 (2012).

\bibitem{MD} M.\ Dutta, S.\ Sarkar, T.\ Ghosh, and D.\ Basak, J.\ Phys.\ Chem.\ C {\bf 116}, 20127 (2012).

\bibitem{Oswald2012} W.\ Oswald and Z.\ Wu, Phys.\ Rev.\ B \textbf{85}, 115431 (2012).
\bibitem{Petersen2011} R.\ Petersen, T.\ G.\ Pedersen, and A.-P.\ Jauho, ACS Nano \textbf{5}, 523 (2011).
\bibitem{JB} J.\ Bai, X.\ Zhong, S.\ Jiang, Y. Huang, and X.\ Duan, Nat.\ Nanotechnol.\ {\bf 5}, 190  (2010).

\bibitem{Maarouf2013} A.\ A.\ Maarouf, R.\ A.\ Nistor, A.\ Afzali-Ardakani, M.\ A.\ Kuroda, D.\ M.\ Newns, and G.\ J.\ Martyna, 
                      J.\ Chem.\ Theory Comput.\ \textbf{9}, 2398 (2013).

\bibitem {DJ} D.\ Jiang, V.\ R.\ Cooper, and S.\ Dai, Nano Lett.\ {\bf 9}, 4019 (2009).

\bibitem{OA} O.\ Akhavan, ACS Nano {\bf 4}, 4174 (2010).
\bibitem{RKP} R.\ K.\ Paul, S.\ Badhulika, N.\ M.\ Saucedo, and A.\ Mulchandani, Anal.\ Chem.\ {\bf 84}, 8171 (2012).
\bibitem{AD} A.\ Du, Z.\ Zhu, and S.\ C.\ Smith, J.\ Am.\ Chem.\ Soc.\ {\bf 132}, 2876 (2010). 
\bibitem{SH} S.-H.\ Huang, L.\ Miao, Y.-J.\ Xiu, M.\ Wen, C.\ Li, L.\ Zhang, and J.-J.\ Jiang, J.\ Appl.\ Phys.\ {\bf 112}, 124312 (2012).
\bibitem{Guo2014} J.\ Guo, J.\ Lee, C.\ I.\ Contescu, N.\ C.\ Gallego, S.\ T.\ Pantelides, S.\ J.\ Pennycook, B.\ A.\ Moyer, and M.\ F.\ Chisholm,
                  Nat.\ Commun.\ \textbf{5}, 5389 (2014).

\bibitem{HXY} H.-X.\ Yang, M.\ Chshiev, D.\ W.\ Boukhvalov, X.\ Waintal, and S.\ Roche, Phys.\ Rev.\ B.\ {\bf 84}, 214404 (2011).
\bibitem{TKa} T.\ Kato, T.\ Nakamura, J.\ Kamijyo, T.\ Kobayashi, Y.\ Yagi, and J.\ Haruyama, Appl.\ Phys.\ Lett.\ {\bf 104}, 252410 (2014).

\bibitem{Lin2015} Y.-C.\ Lin, P.-Y.\ Teng, P.-W.\ Chiu, and K.\ Suenaga, Phys.\ Rev.\ Lett.\ \textbf{115}, 206803 (2015).
\bibitem{Scheffler2006} J.\ M.\ Carlsson and M.\ Scheffler, Phys.\ Rev.\ Lett.\ \textbf{96}, 046806 (2006).
\bibitem{Krashen} A.\ V.\ Krasheninnikov, P.\ O.\ Lehtinen, A.\ S.\ Foster, P.\ Pyykk\"o, and R.\ M.\ Nieminen, Phys.\ Rev.\ Lett. {\bf 102}, 126807 (2009).
\bibitem{Santos2010} E.\ J.\ G.\ Santos, A.\ Ayuela, and D. S\'anchez-Portal, New J.\ Phys.\ {\bf 12}, 053012 (2010).

\bibitem{Fadlallah2016} M.\ M.\ Fadlallah, U.\ Eckern, A.\ H.\ Romero, and U.\ Schwingenschl\"ogl, New J.\ Phys.\ \textbf{18}, 023050 (2016).
              
\bibitem{GK} G.\ Kresse and J.\ Hafner, Phys.\ Rev.\ B {\bf 47}, 558 (1993).
\bibitem{GK1} G.\ Kresse and J.\ Furthm\"uller, Phys.\ Rev.\ B {\bf 54}, 11169 (1996).
\bibitem{JP} J.\ P.\ Perdew, Phys.\ Rev.\ B {\bf 33}, 8822 (1986).
\bibitem{ADB} A.\ D.\ Becke, Phys.\ Rev.\ A {\bf 38}, 3098 (1988).
\bibitem{JP1} J.\ P.\ Perdew, K.\ Burke, and M.\ Ernzerhof, Phys.\ Rev.\ Lett.\ {\bf 77}, 3865 (1996).

\bibitem {HJM} H.\ J.\ Monkhorst and J.\ D.\ Pack, Phys.\ Rev.\ B {\bf 13}, 5188 (1976).
\bibitem{Mavropoulos2004} Ph.\ Mavropoulos, K.\ Sato, R.\ Zeller, P.\ H.\ Dederichs, V.\ Popescu, and H.\ Ebert, Phys.\ Rev.\ B {\bf 69}, 054424 (2004).

\bibitem{Cordero2008} B. Cordero, V. G\'omez, A. E. Platero-Prats, M. Rev\'es, J. Echeverr\'{\i}a, E. Cremades, F. Barrag\'an, and S. Alvarez,
                      Dalton Trans., 2832 (2008). 
\bibitem{Wallace1947} P.\ R.\ Wallace, Phys.\ Rev.\ {\bf 71}, 622 (1947).
\bibitem{Petersen2009} R.\ Petersen and T.\ G.\ Pedersen, Phys.\ Rev.\ B {\bf 80}, 113404  (2009).
\bibitem{Liu2009} W.\ Liu, Z.\ F.\ Wang, Q.\ W.\ Shi, J.\ Yang, and F.\ Liu,  Phys.\ Rev.\ B {\bf 80}, 233405 (2009).

\bibitem{Kokado2006} S.\ Kokado, N.\ Fujima, K.\ Harigaya, H.\ Shimizu, and A.\ Sakuma, Phys.\ Rev.\ B {\bf 73}, 172410 (2006).
\bibitem{Kokado2016} S.\ Kokado, Y.\ Sakuraba, and M.\ Tsunoda, Jpn.\ J.\ Appl.\ Phys.\ {\bf 55}, 108004 (2016).

\end{thebibliography}
\end{document}